\documentclass[aps,prc,twocolumn,balancelastpage,groupedaddress,showpacs]{revtex4-1}
\usepackage{amsmath}
\usepackage{amssymb}
\usepackage{graphicx}
\usepackage{subfig}
\usepackage{caption}
\usepackage{float}
\usepackage{multirow}
\usepackage{array}
\usepackage{bigstrut}
\usepackage[colorlinks=true, citecolor=blue, linkcolor=blue, urlcolor=blue]{hyperref}
\begin{document}

\title{Energy and centrality dependence of particle multiplicity in\\
heavy ion collisions from $\sqrt{s_{_{NN}}}$ = 20 to 2760 GeV}

\author{Leo Zhou}
\email[]{leozhou@mit.edu}
\affiliation{Massachusetts Institute of Technology, Cambridge, Massachusetts 02139, USA}

\author{George S.F. Stephans}
\email[]{gsfs@mit.edu}
\affiliation{Massachusetts Institute of Technology, Cambridge, Massachusetts 02139, USA}

\date{\today}

\begin{abstract}
The centrality dependence of midrapidity charged-particle multiplicities at a nucleon-nucleon center-of-mass energy of 2.76~TeV from CMS are compared to PHOBOS data at 200 and 19.6~GeV.
The results are first fitted with a two-component model which parameterizes the separate contributions of nucleon participants and nucleon-nucleon collisions.
A more direct comparison involves ratios of multiplicity densities per participant pair between the different collision energies.
The results support and extend earlier indications that the influences of centrality and collision energy on midrapidity charged-particle multiplicities are to a large degree independent.
\end{abstract}

\maketitle

\section{Introduction}

With the start of operations at the Large Hadron Collider (LHC), the center-of-mass energies of data for heavy ion collisions now extend up to $\sqrt{s_{_{NN}}}=$ 2.76~TeV.
One of the most basic observables which can be used to understand the hot and dense systems formed in such interactions of ultrarelativistic nuclei is the charged-particle multiplicity density ($dN_{ch}/d\eta$) in the midrapidity region.
The pseudorapidity, defined as $\eta=-\ln[\tan(\theta/2)]$, is used as a substitute for rapidity in these measurements, where $\theta$ is the polar angle with respect to the beam direction.
Furthermore, the dependences of charged-particle densities on centrality (a percentile measure of the impact parameter of the collisions) as well as on collision energy are important in understanding how both ``hard'' and ``soft'' interactions contribute to particle production in the collisions.
Results for $dN_{ch}/d\eta$ in PbPb collisions at $\sqrt{s_{_{NN}}}=$ 2.76 TeV have been extracted at the LHC.
In this paper, these new LHC data are compared with those from AuAu collisions at $\sqrt{s_{_{NN}}}=$~200 and 19.6~GeV using results from the Relativistic Heavy Ion Collider (RHIC).

Similar comparisons of charged-particle multiplicities in heavy ion collisions have been performed using RHIC data from the PHOBOS detector at $\sqrt{s_{_{NN}}}$ values of 19.6, 62.4, 130, and 200~GeV~\cite{Back2004, Alver2009, PhysRevC.74.021901}.
An overview of these and all other PHOBOS multiplicity results can be found in Ref.~\cite{PhobosBigMult}.
Since the centrality dependence was found to be essentially identical within uncertainties for all four RHIC energies, only the two most extreme ones are included in the present analysis.

These earlier analyses used two approaches to study the dependence of $dN_{ch}/d\eta$ on centrality at different collision energies:
\begin{enumerate}
\item Parameterizing the centrality dependence using a two-component model proposed in Ref.~\cite{Kharzeev}:
    \begin{equation} \label{eq:twopara}
    \frac{dN_{ch}}{d\eta}=n_{pp}\Big( (1-x)\frac{\langle N_{part}\rangle}{2} + x \langle N_{coll}\rangle\Big)
    \end{equation}
where $\langle N_{part}\rangle$ and $\langle N_{coll}\rangle$ are respectively the average numbers of participating nucleons and binary nucleon-nucleon collisions in bins of heavy ion interaction centrality.
In this model, the fit parameter $n_{pp}$ represents the average charged-particle multiplicity in a single nucleon-nucleon collision, while $x$ signifies the relative contribution of so-called ``soft" and ``hard" parton interactions.
In heavy ion collisions, ``hard" scatterings, characterized by the production of large transverse momentum ($p_T$), are expected to scale with $N_{coll}$.
In contrast, production of lower $p_T$ particles is expected to scale with $N_{part}$.
The variation of the fit values could then be studied as a function of collision energy.
In a recent study by the PHENIX Collaboration at RHIC, submitted after this present work, the centrality dependence described by Eq.~\eqref{eq:twopara} with $x=0.08$ is found to be very closely equivalent to a linear dependence using a constituent-quark participant model \cite{PHENIX2014}.

\item Computing the pairwise ratios between the multiplicity densities per participant pairs, $dN_{ch}/d\eta/\langle N_{part}/2\rangle$, in corresponding centrality percentile bins at the various energies.
\end{enumerate}
In the second approach, in order to address the fact that $\langle N_{part}\rangle$ is different in respective percentile centrality bins for the two energies, the ratios were calculated in two ways.
One was directly computing the ratios of $dN_{ch}/d\eta/\langle N_{part}/2\rangle$ in the same centrality bin at the two energies but attributing this ratio to the average of the two $\langle N_{part}\rangle$ values.
The other was rebinning the raw collision data at one of the energies with new centrality percentile cutoffs so that $\langle N_{part}\rangle$ was close to the same for the corresponding bins in both data sets.
Reference~\cite{Back2004} found that both types of ratio calculations ``are in agreement, even within the significantly reduced systematic errors.''

In this paper, both the first approach and a slightly modified version of the second approach are employed for the pairwise comparisons of collisions at three different center-of-mass energies from the LHC and RHIC.
It was deemed necessary to modify the second approach because the differences between $\langle N_{part}\rangle$ in corresponding percentile centrality bins at the LHC and RHIC energies are quite large and rebinning to accommodate matching $\langle N_{part}\rangle$ would require significant reanalysis of the raw data sets.
Instead, a more efficient method was chosen, in which multiplicity ratios were calculated by linearly interpolating the $dN_{ch}/d\eta/\langle N_{part}/2\rangle$ at one energy to the $\langle N_{part}\rangle$ of each centrality bin in the other data set.

\section{Method of Analysis\label{sec:data}}
\subsection{Data Source}
RHIC data for midrapidity charged-particle multiplicity have been presented by the BRAHMS~\cite{BRAHMS}, PHENIX~\cite{PHENIX}, PHOBOS~\cite{PHOBOSWhite, PhobosBigMult}, and STAR~\cite{STAR} Collaborations.
Since the PHOBOS Collaboration provides the most extensive set of published multiplicity results, it was chosen for this analysis over the other RHIC experiments.
The \mbox{PHOBOS} data for charged-particle multiplicities in AuAu collisions at 19.6 and 200 GeV are taken from Refs.~\cite{Back2004} and \cite{Alver2009}.
These PHOBOS midrapidity charged-particle multiplicity data are averaged over the pseudorapidity region $|\eta|<1$.
In addition, Ref.~\cite{Back2004} lists the  charged-particle multiplicity data for inelastic $p(\bar{p})+p$ collisions at 200 GeV ($dN_{ch}/d\eta=2.29\pm0.08$) and an interpolated value at 19.6 GeV ($dN_{ch}/d\eta=1.27\pm0.13$).
The $dN_{ch}/d\eta$ for 200 GeV $pp$ collisions directly measured by PHOBOS \cite{PhobosBigMult} is consistent with 2.29 but has a larger uncertainty.

The PbPb collision data at 2.76~TeV from CMS are taken from Ref.~\cite{CMS2.76}.
While the ALICE~\cite{ALICE2.76,ALICE2.76-2} and \mbox{ATLAS}~\cite{ATLAS2.76} Collaborations at the LHC also have heavy ion collision data at the same energy, only CMS data was used in this comparison with lower energy results from PHOBOS.
The CMS heavy ion data extend to more peripheral events, and the $\langle N_{part}\rangle$ values from CMS were extracted with a procedure  similar to that used by \mbox{PHOBOS}.
Furthermore, the ALICE, CMS, and ATLAS heavy ion data for charged-particle multiplicity are consistent over the centrality range where they overlap \cite{ATLAS2.76}, and so the conclusions presented here do not depend on the choice of LHC data set.
Notably, the qualitative similarity in the centrality dependence of
multiplicity data from the LHC and the highest RHIC energy was first mentioned by Ref.~\cite{ALICE2.76-2}.
Both the \mbox{ALICE} and CMS $pp$ data (discussed below), as well as the CMS PbPb data were averaged over $|\eta|<0.5$.

There are no published charged-particle multiplicity data from any experiment for inelastic $pp$ collisions at 2.76~TeV.
The $pp$ data from CMS with the closest energy are the measured non-single-diffractive (NSD) $dN_{ch}/d\eta=4.47\pm0.04(\textrm{stat.})\pm0.16(\textrm{syst.})$ at 2.36~TeV~\cite{CMS2.36}.
At that same lower energy, the ALICE Collaboration found that $dN_{ch}/d\eta=3.77\pm0.01(\textrm{stat.})^{+0.25}_{-0.12}(\textrm{syst.})$ for inelastic $pp$ collisions, and $dN_{ch}/d\eta=4.43\pm0.01(\textrm{stat.})^{+0.17}_{-0.12}(\textrm{syst.})$ for NSD $pp$ collisions~\cite{ALICE2.36}.
These three data points were used to estimate a 2.36~TeV inelastic $pp$ value from CMS, which was then extrapolated to 2.76~TeV using the $\sqrt{s_{_{NN}}}$ dependence of charged-particle multiplicity density given in Ref.~\cite{CMS7}.
Since the ALICE and CMS multiplicities for NSD events agree to better than 1\% and the extrapolation from 2.36 to 2.76~TeV is an increase of just over 3\%, this procedure for getting a comparable inelastic $pp$ multiplicity value at 2.76~TeV does not introduce significant additional uncertainty.

The reported uncertainties in $dN_{ch}/d\eta/\langle N_{part}/2\rangle$ data contain a combination of ``slope" and ``scale" uncertainties added in quadrature.
The scale uncertainty is directly proportional to the value of $dN_{ch}/d\eta/\langle N_{part}/2\rangle$, whereas the slope uncertainty accounts for variations of individual data points.
The latter uncertainty would predominantly shift all points by an amount that varies smoothly with centrality, although there is a small component that varies point to point.
For the fits to Eq.~\eqref{eq:twopara}, scale uncertainties enter into the systematic uncertainty in $n_{pp}$ but have essentially no impact on the uncertainty of $x$.
For the ratio comparison, the scale uncertainties can be ignored when studying the shape of the centrality dependence.

\subsection{Two-Component Fit Comparison}
In addition to $\langle N_{part}\rangle$, values of $\langle N_{coll}\rangle$ are required for comparison of PHOBOS and CMS using the two-component model given by Eq.~\eqref{eq:twopara}.
Reference~\cite{Back2004} used the parameterization $N_{coll}=A\times N_{part}^\alpha$ (found by fitting the results of a Glauber model calculation) with $(A,\alpha)=(0.37, 1.32)$ and $(0.33, 1.37)$ for 19.6 and 200 GeV, respectively.
The $\langle N_{coll}\rangle$ values for 2.76~TeV were the same as those used by CMS~\cite{CMS2.76}.

The systematic uncertainties in the fit parameters, $x$ (the fraction representing the relative contribution from ``hard processes'') and $n_{pp}$ (the average multiplicity from $pp$ collisions), are estimated by reperforming the fit with the uncertainties added to or subtracted from the data points and looking at the resulting variations in fit results.
The first fit used the entire available data set at each energy.
In order to make the most direct possible comparison of fit results, a second fit was performed at 200~GeV and 2.76~TeV with an $N_{part}$ range restricted to that available in the 19.6 GeV data set.

\begin{figure*}[htb]
\includegraphics[width=0.497\textwidth]{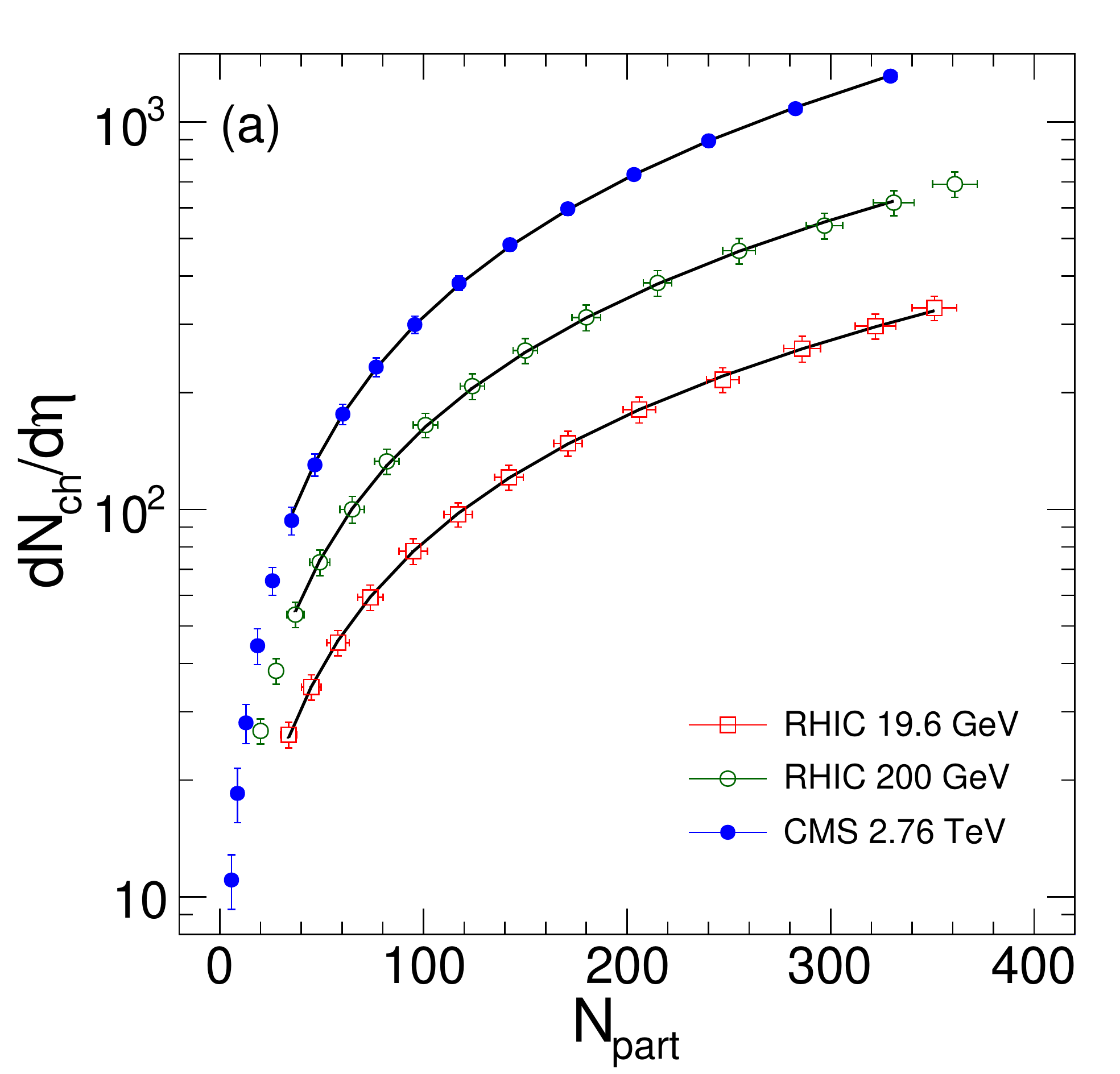}
\includegraphics[width=0.497\textwidth]{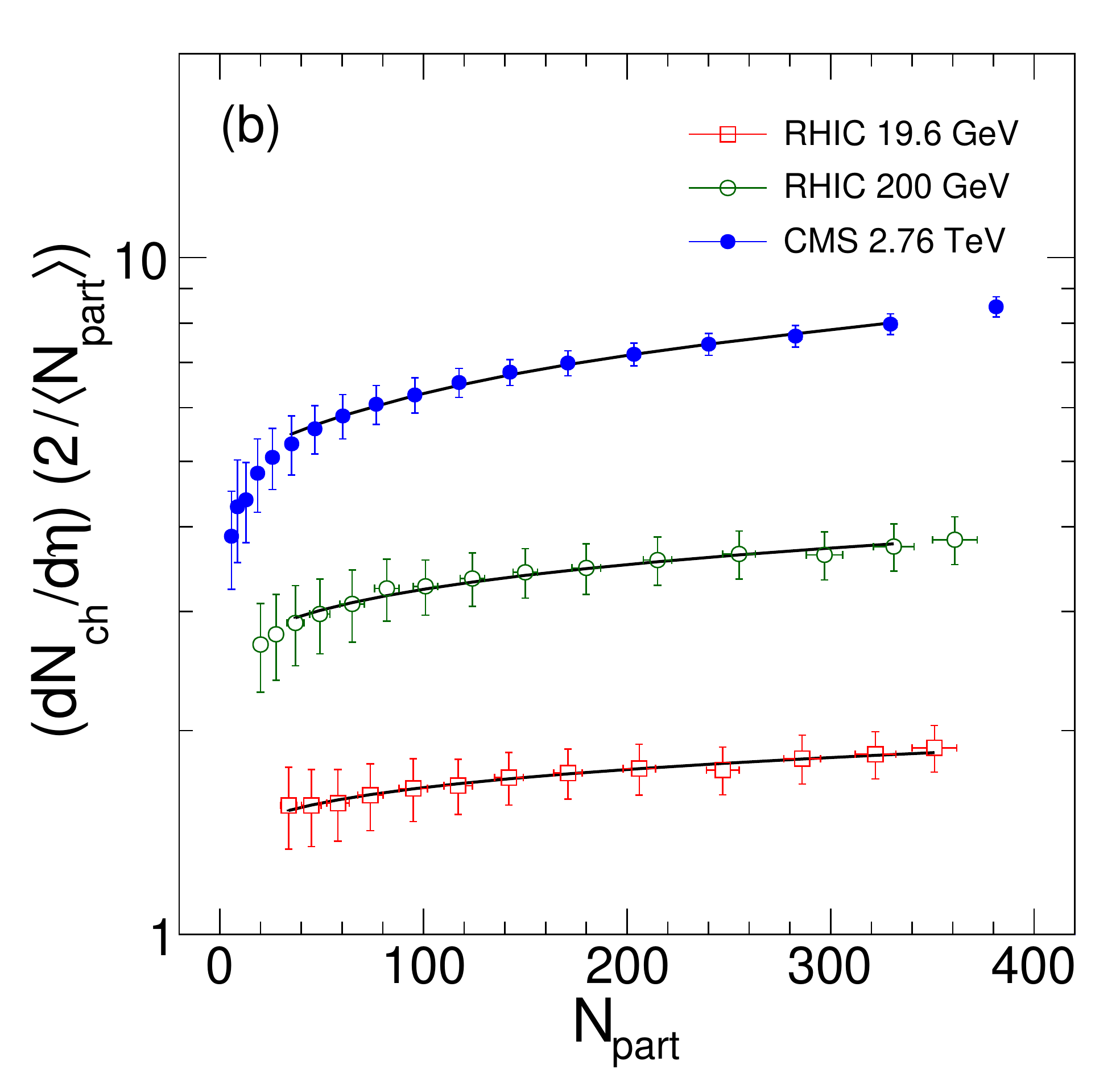}
\caption{\label{fits}(Color online)~ (a) Midrapidity charged-particle multiplicity density $dN_{ch}/d\eta$. (b) $dN_{ch}/d\eta$ divided by the number of participant pairs. In both panels, vertical and horizontal bars show systematic uncertainties in $dN_{ch}/d\eta$ and $\langle N_{part}\rangle$, respectively. Lines show the results of the two-component fit using Eq.~\eqref{eq:twopara}. Only the fits restricted to the common range of $\langle N_{part}\rangle$ are shown.}
\end{figure*}

\begin{table*}[htb]
\centering
\begin{tabular}{l|c|c|c|c}
\hline \hline
 & $n_{pp}$ & $x$ & $pp$ $dN_{ch}/d\eta$ & Data/$n_{pp}$\\ \hline
PHOBOS 19.6 GeV & 1.357 $\pm$ 0.194 & 0.096 $\pm$ 0.045 & 1.27 $\pm$ 0.13 & 0.94 $\pm$ 0.17 \\ \hline
PHOBOS 200 GeV & 2.454 $\pm$ 0.343 & 0.118 $\pm$ 0.042 & \multirow{2}{*}{2.29 $\pm$ 0.08} & 0.93 $\pm$ 0.17\\ \cline{1-3}\cline{5-5}
PHOBOS 200 GeV (restricted $N_{part}$) & 2.531 $\pm$ 0.342 & 0.106 $\pm$ 0.039 & & 0.90 $\pm$ 0.13\\ \hline
CMS 2.76 TeV & 4.373 $\pm$ 0.473 & 0.123 $\pm$ 0.029 & \multirow{2}{*}{3.93 $^{+0.53}_{-0.40}$} & 0.90 $\pm$ 0.14\\ \cline{1-3}\cline{5-5}
CMS 2.76 TeV (restricted $N_{part}$) & 4.686 $\pm$ 0.418 & 0.102 $\pm$ 0.022 &  & 0.84 $\pm$ 0.12\\
\hline \hline
\end{tabular}
\caption{\label{twocompres}Two-component model fit results, reported (or estimated) inelastic $pp$ midrapidity $dN_{ch}/d\eta$, and the ratio of $pp$ $dN_{ch}/d\eta$ data over the fitted value of $n_{pp}$. All uncertainties are systematic.}
\end{table*}

\subsection{Ratio Comparison}
In order to extract direct ratios of multiplicities, it is first necessary to interpolate the data for one of the energies to values of $N_{part}$ which match those for the other energy.
More specifically, the interpolations were performed on the values of $dN_{ch}/d\eta/\langle N_{part}/2\rangle$, since those vary much more slowly with centrality.
For the current analysis, the CMS data at 2.76~TeV were interpolated to the $\langle N_{part}\rangle$ values for the PHOBOS 200~GeV data.
For every PHOBOS point, the two CMS points with the nearest $\langle N_{part}\rangle$ (one higher and one lower) were linearly interpolated to the value of $\langle N_{part}\rangle$ for the PHOBOS point.
Interpolating using more points and a higher order polynomial made essentially no difference in the final results.
The uncertainty of each interpolated point is calculated by also linearly interpolating the fractional slope uncertainties of its two nearest neighboring CMS data points.
Scale uncertainties are ignored for the reasons discussed above.
This interpolation method depends on the facts that the reported uncertainties are entirely systematic and that the fractional slope uncertainties in $dN_{ch}/d\eta/\langle N_{part}/2\rangle$ vary slowly as a function of centrality.
After all the CMS data are properly interpolated, the ratios $R_\text{2760/200}$ are calculated by dividing the corresponding $dN_{ch}/d\eta/\langle N_{part}/2\rangle$ values at each $\langle N_{part}\rangle$.
Uncertainties in these ratios are calculated by adding the fractional uncertainties in quadrature.

The same procedure was used to interpolate the PHOBOS 200~GeV data in order to take ratios with the 19.6~GeV points.
These newly extracted ratios ($R_\text{200/19.6}$) were compared to those calculated using the methods described in Ref.~\cite{Back2004} and were found to be consistent.
The systematic uncertainties in the previous comparisons of different PHOBOS energies were reduced by the fact that some uncertainties were common to all energies and could be factored out.
A detailed evaluation of the methods used by CMS and PHOBOS did not reveal any similar commonalities which could be eliminated.
Therefore, for consistency of presentation, the (larger) systematic uncertainties given by the current analysis procedure were used in comparing ratios of multiplicities for the various energies.
Nevertheless, the difference in uncertainties of $R_{200/19.6}$ between what is presented here and those in Ref.~\cite{Back2004} were reduced by factoring out the scale uncertainties.

\section{Results\label{sec:results}}

Results for the two-component fits of multiplicity for the PHOBOS and CMS data are summarized in Table~\ref{twocompres} and Fig.~\ref{fits}.
The two panels of Fig.~\ref{fits} show the same data and fits, but the multiplicities are divided by the number of participant pairs in the right panel to more easily compare the shapes at the various energies.
In order to make the most meaningful comparison, only the fits restricted to the common range of $\langle N_{part}\rangle$ are shown.
In addition to the fit parameters, Table~\ref{twocompres} lists published (or estimated) midrapidity charged-particle multiplicities for inelastic $pp$ collisions and the ratios of those data to the fitted values of $n_{pp}$.

The two-component fits yield very similar values for the $x$ parameter at all three energies, especially for the fits restricted to a common $\langle N_{part}\rangle$ range.
This suggests that collision energy does not affect the division of particle production between hard and soft scattering, even over an energy range of more than two orders of magnitude.
This observation is in contrast to the expectation that hard scatterings, which scale with the number of nucleon-nucleon collisions, should become increasingly prevalent as the collision energy increases.

It is important to note in this comparison that the relative increases in $\langle N_{part}\rangle$ and $\langle N_{coll}\rangle$ with increasing center-of-mass energy are dramatically different.
For the most central collisions, the values of $\langle N_{part}\rangle$ at 200~GeV and 2.76~TeV are roughly 3\% and 8.5\% larger, respectively, than that at 19.6~GeV.
In contrast, the corresponding values of $\langle N_{coll}\rangle$ are 24\% and almost a factor of 2 larger.
As a result, the relative inputs from the two terms in Eq.~\eqref{eq:twopara} towards the total multiplicity differ somewhat  at the different energies even though the fitted value of $x$ is essentially unchanged.
The stability of $x$ indicates a surprising constancy of the relative contributions of each individual participant pair and each individual nucleon-nucleon collision, even when the collisions overall account for a greater proportion of the total multiplicity at higher energy.

As expected, $n_{pp}$ increases significantly with increasing collision energy.
Table~\ref{twocompres} compares these fit parameters to published (or estimated) values of midrapidity charged-particle multiplicities for inelastic $pp$ collisions at each energy.
As shown in the last column of the table, the fitted values of $n_{pp}$ for all three energies are slightly above the actual $pp$ multiplicities.
Furthermore, the difference between the data and the fit parameters shows almost no energy dependence, as all of the ratios are equal within their systematic uncertainties.
Although there appears to be a decrease in the ratio with increasing beam energy, this trend is quite weak and much less than the systematic uncertainties in the individual points.
Because many of the systematic uncertainties are common between the two fits at 200 and 2760 GeV, the trend for $n_{pp}$ to approach the $pp$ multiplicity value as a larger range of $N_{part}$ is included may be more significant.

\begin{figure}[htb]
\includegraphics[width=0.48\textwidth]{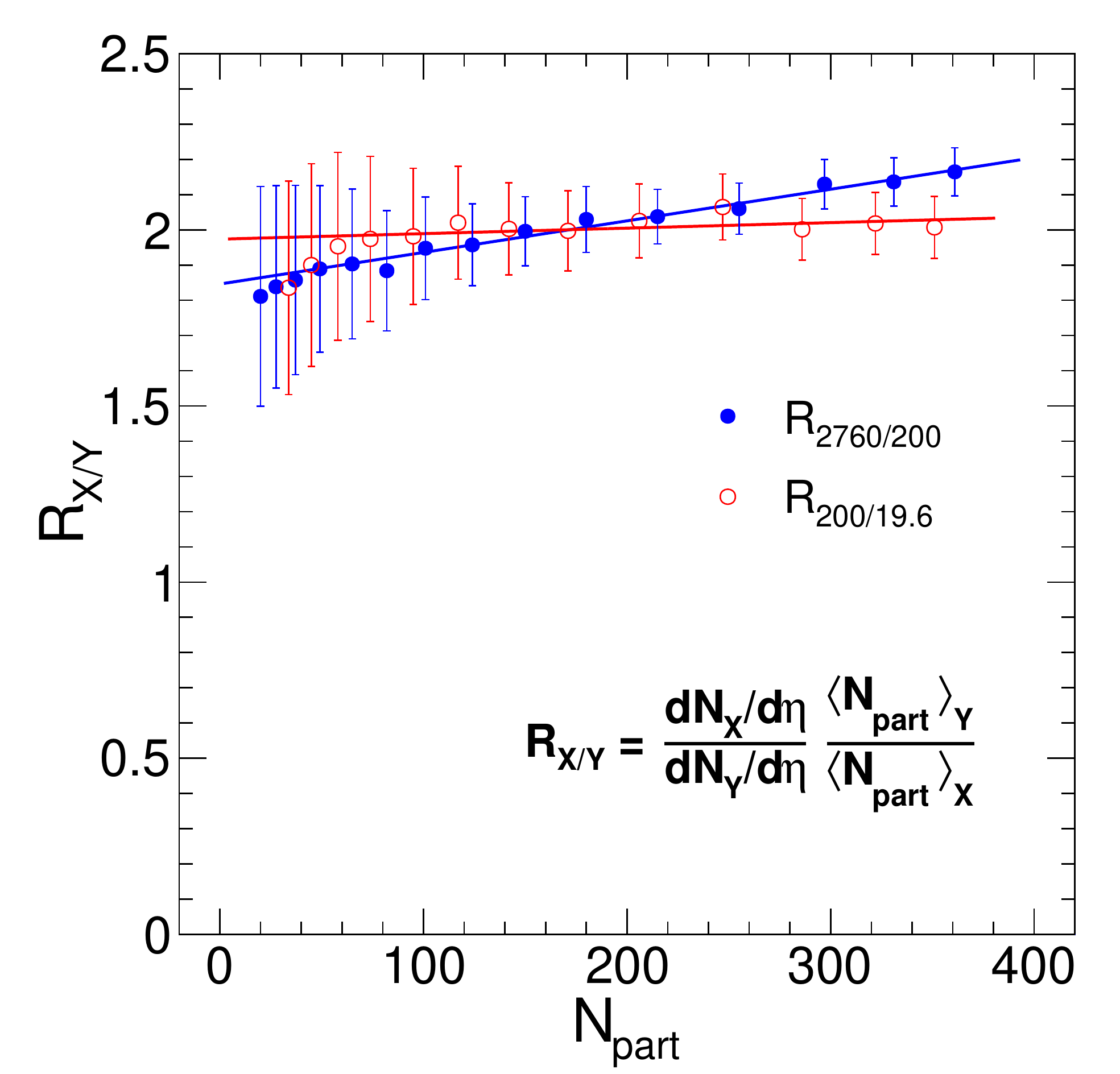}
\caption{(Color online)~ Double ratios of midrapidity $dN_{ch}/d\eta$/$\langle N_{part}\rangle$. The lines are linear fits to guide the eye. Error bars show only the slope uncertainties. The scale uncertainties are 8.3\% for $R_\text{2760/200}$ and 10.7\% for $R_{200/19.6}$.}
\label{ratios}
\end{figure}

For a less model-dependent comparison, the multiplicity ratios $R_\text{2760/200}$ and $R_\text{200/19.6}$ (where the higher energy data are interpolated to the same $N_{part}$ values as those for the lower energy) are plotted versus centrality in Fig.~\ref{ratios}.
The combined slope uncertainties would predominantly shift all points of a given set of ratios in the same direction by a varying amount proportional to the size of the error bars shown.
The scale uncertainties (not shown) would move all points together by the same multiplicative factor.

It is entirely coincidental that the average values of these two ratios are almost identical in magnitude.
The more significant fact is that the two ratios are very similar in shape.
This is consistent with what was observed in Refs.~\cite{Back2004, Alver2009, PhysRevC.74.021901} where multiplicity ratios over a narrower range of collision energies showed little or no centrality dependence.
While there is some indication that the centrality dependence at the highest energy might be slightly steeper, the slopes from linear fits of both ratios versus $N_{part}$ are consistent with each other within the uncertainties, and both slopes are close to zero.
Thus, this expanded analysis supports and extends the claim that midrapidity charged-particle multiplicities ``factor'' into independent energy and centrality dependences.

This factorization has also been explored at RHIC energies by fitting all of the PHOBOS data using separate empirical functions of energy and $N_{part}$ alone [i.e. not as a function of both $N_{part}$ and $N_{coll}$ as in Eq.\eqref{eq:twopara}]~\cite{PhobosBigMult}.
However, this fitted centrality dependence for RHIC results is a very poor match to the CMS data, probably because the relationship between $N_{part}$ and $N_{coll}$ is very different at the LHC energy.

\section{Conclusion\label{sec:conclusion}}
Midrapidity charged-particle multiplicity data for heavy ion collisions at three broadly spaced center-of-mass energies were compared using two different approaches.
Both analyses show that the energy and centrality dependences of the multiplicity are largely independent.
Increasing the collision energy does not appear to alter the shape of the centrality dependence of multiplicity.
In particular, the fractional contribution of ``hard processes'', as parameterized by the value of $x$ in the two-component model, does not vary significantly, even over a range of more than two orders of magnitude in collision energy.
In a more direct comparison, ratios of midrapidity multiplicity densities show at most a very weak dependence on collision energy or centrality.
This rough ``factoring'' of the centrality and energy dependencies of midrapidity charged-particle multiplicity, as first proposed in Ref.~\cite{Back2004} using PHOBOS data spanning $\sqrt{s_{_{NN}}}=19.6$ to $200$~GeV, is found to extend largely unaltered up to $\sqrt{s_{_{NN}}}=2760$~GeV.

This work was partially supported by U.S. DOE Grant No.~DE-FG02-94ER40818 and by the Massachusetts Institute of Technology Undergraduate Research Opportunities Program.

\bibliography{PHOBOSCMS}
\end{document}